\documentclass{article}

\usepackage{arxiv}

\usepackage[utf8]{inputenc} 
\usepackage[T1]{fontenc}    
\usepackage{hyperref}       
\usepackage{url}            
\usepackage{booktabs}       
\usepackage{amsfonts}       
\usepackage{nicefrac}       
\usepackage{microtype}      
\usepackage{lipsum}		
\usepackage{graphicx}

\usepackage[square,sort,comma,numbers]{natbib}

\usepackage{doi}

\usepackage{caption}
\usepackage{subcaption}

\usepackage{algorithm} 
\usepackage[noend]{algpseudocode}

\newcommand{\blankline}{\vspace{\baselineskip} }

\title{Spontaneous Emergence of Computation in Network Cascades}

\author{ Galen J. Wilkerson\\
Centre for Complexity Science and\\ Department of Mathematics\\
Imperial College London\\
South Kensington Campus, SW7 2AZ, UK\\
and\\
Department of Computer Science\\
University of Surrey\\
Guildford, GU2 7XH, UK\\
\AND
Sotiris Moschoyiannis\\
Department of Computer Science\\
University of Surrey\\
Guildford, GU2 7XH, UK\\
\AND
Henrik Jeldtoft Jensen\\
Centre for Complexity Science and\\ Department of Mathematics\\
Imperial College London\\
South Kensington Campus, SW7 2AZ, UK\\
and\\
Institute of Innovative Research\\ Tokyo Institute of Technology\\ Nagatsuta-cho, Yokohama 226-8502, Japan\\
}

\renewcommand{\shorttitle}{Spontaneous Emergence of Computation in Network Cascades}

\hypersetup{
pdftitle={Spontaneous Emergence of Computation in Network Cascades},
pdfsubject={cs.AI, cs.LO, cs.NE, cs.CC, cs.FL, cs.LG, cs.MA, math.CO, math.IT, math.LO, math.MP, math.PR, math.ST, cond-mat.dis-nn, cond-mat.stat-mech, math-ph, nlin.AO, physics.soc-ph, q-bio.NC, q-fin.CP, stat.AP, stat.CO, stat.ML, stat.TH},
pdfauthor={Galen Wilkerson, Sotiris Moschoyiannis, Henrik Jeldtoft Jensen},
pdfkeywords={network cascades, computation, Boolean logic, information processing,  computational neuroscience, deep learning, criticality, percolation, complexity},
}

\begin{document}
\maketitle

\begin{abstract}
Neuronal network computation and computation by avalanche supporting networks are of interest to the fields of physics, computer science (computation theory as well as statistical or machine learning) and neuroscience.  Here we show that computation of complex Boolean functions arises spontaneously in threshold networks as a function of connectivity and antagonism (inhibition), computed by \textit{logic automata (motifs)} in the form of \textit{computational cascades}.  We explain the emergent inverse relationship between the computational complexity of the motifs and their rank-ordering by function probabilities due to motifs, and its relationship to symmetry in function space.  We also show that the optimal fraction of inhibition observed here supports results in computational neuroscience, relating to optimal information processing.\end{abstract}

\keywords{network cascades\and computation \and Boolean logic \and information processing \and computational neuroscience \and deep learning \and criticality \and percolation \and complexity}

\newpage

\section{Introduction}

The relationship between physical systems and information has been of increasing and compelling interest in the domains of physics \cite{jaynes1957information,wheeler1992recent,ben2008farewell}, neuroscience \cite{sejnowski1988computational, lynn2019physics}, computer science \cite{shannon1941mathematical, turing1936computable, dubbey2004mathematical,wolfram2002new,conway1970game,lizier2014framework, rojas2013neural}, quantum computing \cite{lloyd2013universe, wheeler1992recent,perseguers2010quantum}, and other fields such as computation in social networks \cite{schelling1971dynamic, granovetter1978threshold, sakoda1971checkerboard}, or biology \cite{schrodinger1992life,brooks1988evolution} to the point where some consider information to be a fundamental phenomenon in the universe \cite{lloyd2010computational,wheeler2018information,knuth2011information}.  Often, physical systems operating on information take place on, or can be modeled by, network activity \cite{watts2002simple}, since information is transmitted and processed by interactions between physical entities.

The principle of Occam’s razor and goals of achieving a deeper understanding of these physical-information interactions encourage us to find the simplest possible processes achieving computation.  Thus we may conduct \textit{basic research into understanding necessary and sufficient conditions for systems to perform information processing}.

Cascades, particularly on networks, are such a simple and ubiquitous process.  Cascades are found in a great number of systems – the brain, social networks, chemical-, physical- , and biological- systems – occurring as neuronal avalanches, information diffusion, influence spreading, chemical reactions, chain reactions, activity in granular media, forest fires or metabolic activity, to name a few \cite{watts2002simple,kempe2003maximizing,newman2018networks, easley2010networks,christensen2005complexity,jalili2017information}.  
The Linear Threshold Model (LTM) is among the simplest theoretical models to undergo cascades.  As a simple threshold network, the LTM is also similar to artificial models of neural networks, without topology restrictions \cite{rojas2013neural}.


Since the work of Shannon \cite{shannon1948mathematical}, the \textit{bit} has been considered the basic unit of information.  Therefore, whatever we can learn about processing of bits can be extended to information processing in non-Boolean systems.  The tools of Boolean logic then allow us to begin to develop a formalism linking LTM and other cascades to information processing in the theory of computing \cite{von1956probabilistic}.   In systems of computation or statistical learning, patterns of inputs are mapped to patterns of output by Boolean functions \cite{savage1998models, rojas2013neural}.

Another way to express this is that a bit is the simplest possible perturbation of a system.  Bits can interact via some medium, these interactions can be represented by edges in a network, and Boolean functions describe the results of possible interaction patterns.

Since we aim to study this topic from first principles, we are interested in how the combinatorial space of possible networks interacts with the combinatorial space of possible Boolean functions, via cascades and the control parameters.   Particularly, we would like to understand the \textit{phase space of Boolean functions} computed by LTM nodes on the input (seed) nodes by the cascade action.  

From a mathematical perspective, we can treat the brain or other natural systems having $N$ elements in the worst case as a random network, where there are $N(N-1)/2$ possible connections, yielding $2^{(N(N-1)/2)}$ possible networks.  Meanwhile, the space of Boolean functions grows exceptionally quickly.  There are $2^{2^k}$ unique Boolean functions on $k$ inputs.   This immediately makes us ask how this space behaves, and how large networks such as the brain can navigate toward particular functions in this vast space.
We also observe that for the all the functions available on $k$ inputs, the decision tree complexity (depth of decision tree computing them) appears exponentially distributed, meaning that the vast majority of functions available are complex as $k$ increases.

A somewhat surprising initial result in this investigation is that \textit{complex functions on inputs emerge spontaneously and seemingly inevitably as threshold networks are connected at random}.

\section{Linear Threshold Model (LTM), Boolean logic, and antagonism}

The Linear Threshold model (LTM) \cite{watts2002simple} is defined as follows:  A random (Erdos-Renyi-Gilbert) graph is constructed, having $N$ nodes and $p$, the probability of an edge between each pair of nodes.  Each node is then assigned a random threshold $\phi$ from a uniform distribution, $\phi \sim U[0,1]$.  Nodes can be \textit{unlabelled} or \textit{labelled}, and are all initialized as \textit{unlabelled}.  To run the cascade, a small set of seed nodes are \textit{perturbed}, marked as labelled. Now, each unlabelled node $u$ is examined randomly and asynchronously, and the fraction of its graph neighbors that are labelled $\big(\frac{L(u)}{deg(u)}\big)$ is determined, where $L(u)$ is the number of $u$'s neighbors that are labeled, and $deg(u)$ is $u$'s degree.  If $u$'s fraction reaches its threshold $\big(\frac{L(u)}{deg(u)}\ge \phi \big)$, $u$ is marked labelled.  This process continues until no more nodes become labelled.  Here we note that the LTM may be written in vector form, and bears some similarity to the artificial McCulloch-Pitts neuron \cite{rojas2013neural}.


It has been shown that the LTM exhibits \textit{percolation}, where a giant connected component (GCC) of easily-influenced \textit{vulnerable} nodes $u$ (having $\phi \le 1/deg(u)$, where $deg$ is the degree of $u$) suddenly arises at the critical connectivity \cite{watts2002simple}.  

We observe that cascades in the LTM compute monotone Boolean functions (the number of true outputs cannot decrease in the number of true inputs) at each node on input perturbation patterns \cite{wilkerson2019universal}.  In our numerical experiments, we create the LTM as above, but choose input seed nodes $a$ and $b$ (for $k = 2$ inputs) as the only possible loci of initial perturbation.  In one trial, we create a network, freezing network edges and thresholds across all possible input patterns [Table \ref{truth_table2}, cols. a, b]. For each input pattern we reset non-seed nodes to unlabelled, set seeds according to inputs, and run the cascade.  We then identify the function computed by each node ($f_0, ..., f_{15}$) [Table \ref{truth_table2}, cols. 0-15].

\begin{table}[htbp]
\centering
\begin{tabular}{|c|c||c|c|c|c|c|c|c|c|c|c|c|c|c|c|c|c|}
\hline

\multicolumn{2}{|c||}{inputs}&\multicolumn{16}{c|}{functions}  \\ \hline
a  & b & 0 & 1& 2& 3& 4& 5& 6& 7& 8& 9& 10& 11& 12& 13& 14& 15 \\ \hline
0 & 0 & 0& 0& 0& 0& 0& 0& 0& 0& 1& 1& 1& 1& 1& 1& 1& 1\\
0 & 1 &  0& 0&  0& 0&  1& 1& 1& 1& 0& 0&  0& 0&  1& 1& 1& 1\\
1 & 0 & 0& 0& 1& 1&0& 0& 1& 1&0& 0& 1& 1&0& 0& 1& 1\\
1 & 1 & 0 & 1 & 0 & 1 & 0 & 1 & 0 & 1 & 0 & 1 & 0 & 1 & 0 & 1 & 0 & 1 \\
\hline
\end{tabular}
\blankline

\caption{Truth tables for binary functions.  The truth tables of all possible unique binary $(k =2)$ Boolean functions are shown ($2^{2^k} = 16$ functions).  The LTM can only compute \textit{monotonically-increasing} Boolean functions (columns 0, 1, 3, 5, 7), where the first row equals zero, since the seed nodes are unlabelled.  Thus, it cannot compute functions 2, 4, 6, or 8 to 15.}
\label{truth_table2}
\end{table}

\begin{figure}[ht]
    \centering
    \includegraphics[scale = 0.7]{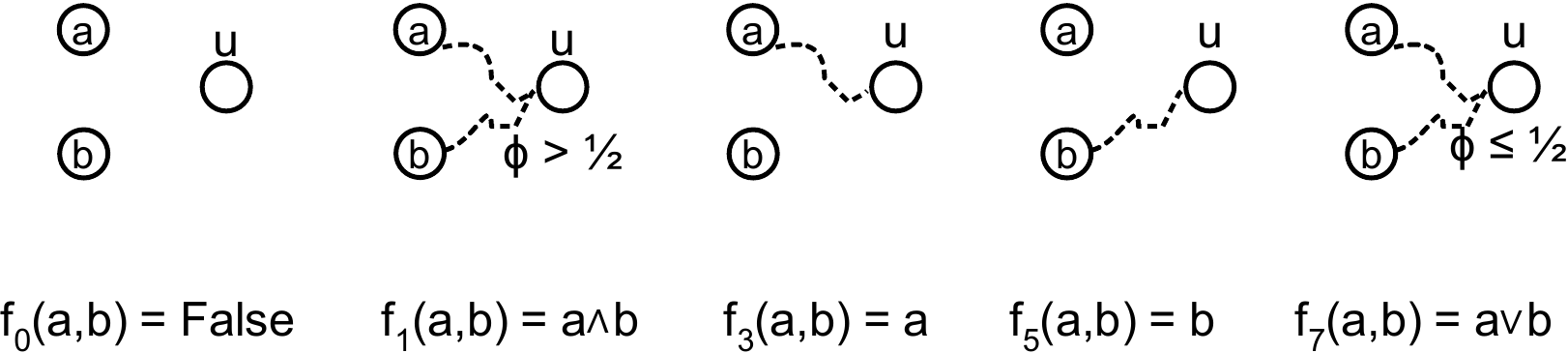}
    \caption{\textit{Logic motifs} compute Boolean functions.  The simplest LTM sub-networks are logical automata \textit{(logic motifs)} and compute the monotone functions for $k = 2$ inputs at node $u$ on perturbations of $a$ and $b$.  Dashed lines are network paths.}
    \label{fig:simplest_networks}
\end{figure}

 The zero function, $f_0(a,b) = 0$ (False) is computed by a simple sub-network, where node $u$ has no path to either seed node [Fig. \ref{fig:simplest_networks}].  Similarly, function $f_1(a,b) = a \wedge b$ (AND) is computed by $u$ with a sub-network having paths from both seed nodes $a,b$, and a threshold $\phi > \frac 12$.  Similar sub-networks allow us to obtain nodes computing monotone functions $f_3, f_5, f_7$ [Fig. \ref{fig:simplest_networks}].  These sub-networks are therefore logical automata \cite{von1951general,von1956probabilistic}, and we note that they form functional \textit{logic motifs} in the network \cite{milo2002network}.

\begin{figure}[ht]
    \centering
          \includegraphics[scale = 0.70]{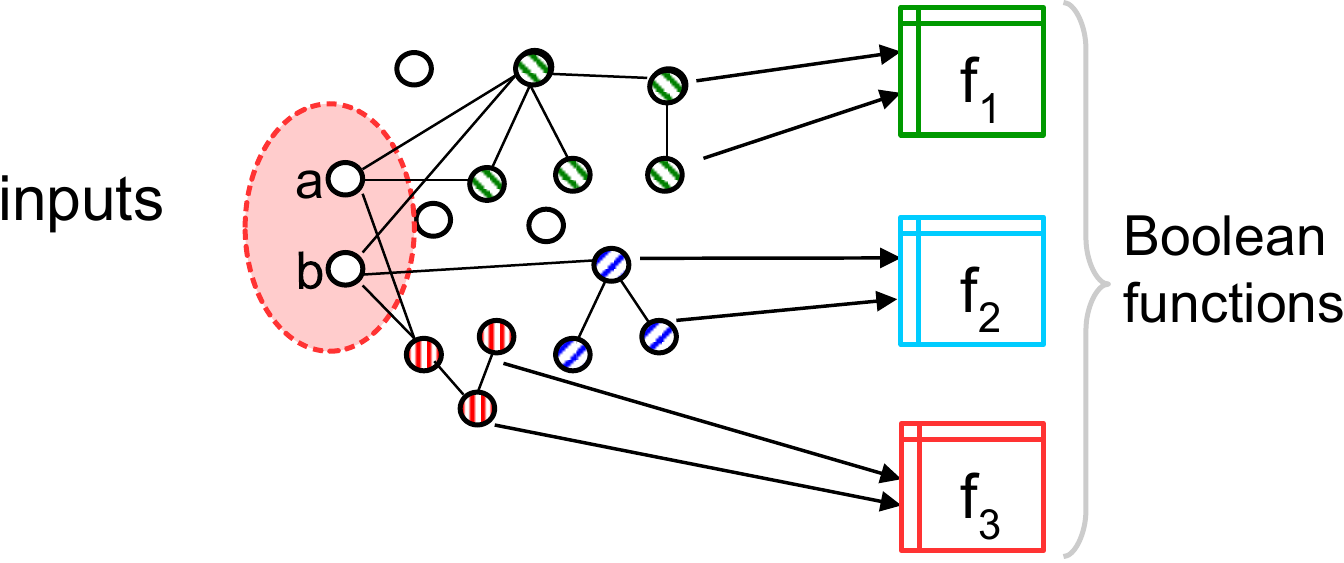}
        \caption{LTM nodes compute Boolean functions in \textit{computational cascades}.  Iterating through all possible perturbations of input seed nodes $a$ and $b$, each network node must compute some Boolean function on the inputs.}
        \label{fig:schematic}
\end{figure}

We find that an LTM network cascade will yield a distribution of Boolean functions on its input nodes, and the possible functions computed by network nodes will partition the set of monotone Boolean functions [Fig. \ref{fig:schematic}] (with the exception of $f_{15}$).
Thus the LTM carries out \textit{computational cascades} on input perturbation patterns.

We then obtain monotonically decreasing functions  (negation of the LTM), by taking the logical complement of the original LTM labelling rule, so that some node $u$ is instead activated when its \textit{fraction of labelled neighbors is less than its threshold $\big(\frac{L(u)}{deg(u)} < \phi\big)$}.  We call such nodes \textit{antagonistic}, from which we can construct an \textit{antagonistic linear threshold model (ALTM)}.  For 2 inputs, replacing $u$ with an ALTM node $\neg u$, will compute $f_{15}, f_{14}, f_{12}, f_{10}$, and $f_8$ [Table \ref{truth_table2}], and the sub-networks are antagonistic versions of those for $f_0, f_1, f_3, f_5,$ and $f_7$, respectively [Fig. \ref{fig:simplest_networks}].
\begin{figure}[ht]
     \centering{
     \begin{subfigure}[t]{0.47\textwidth}         
     \vskip 0pt
     \includegraphics[scale =.42]{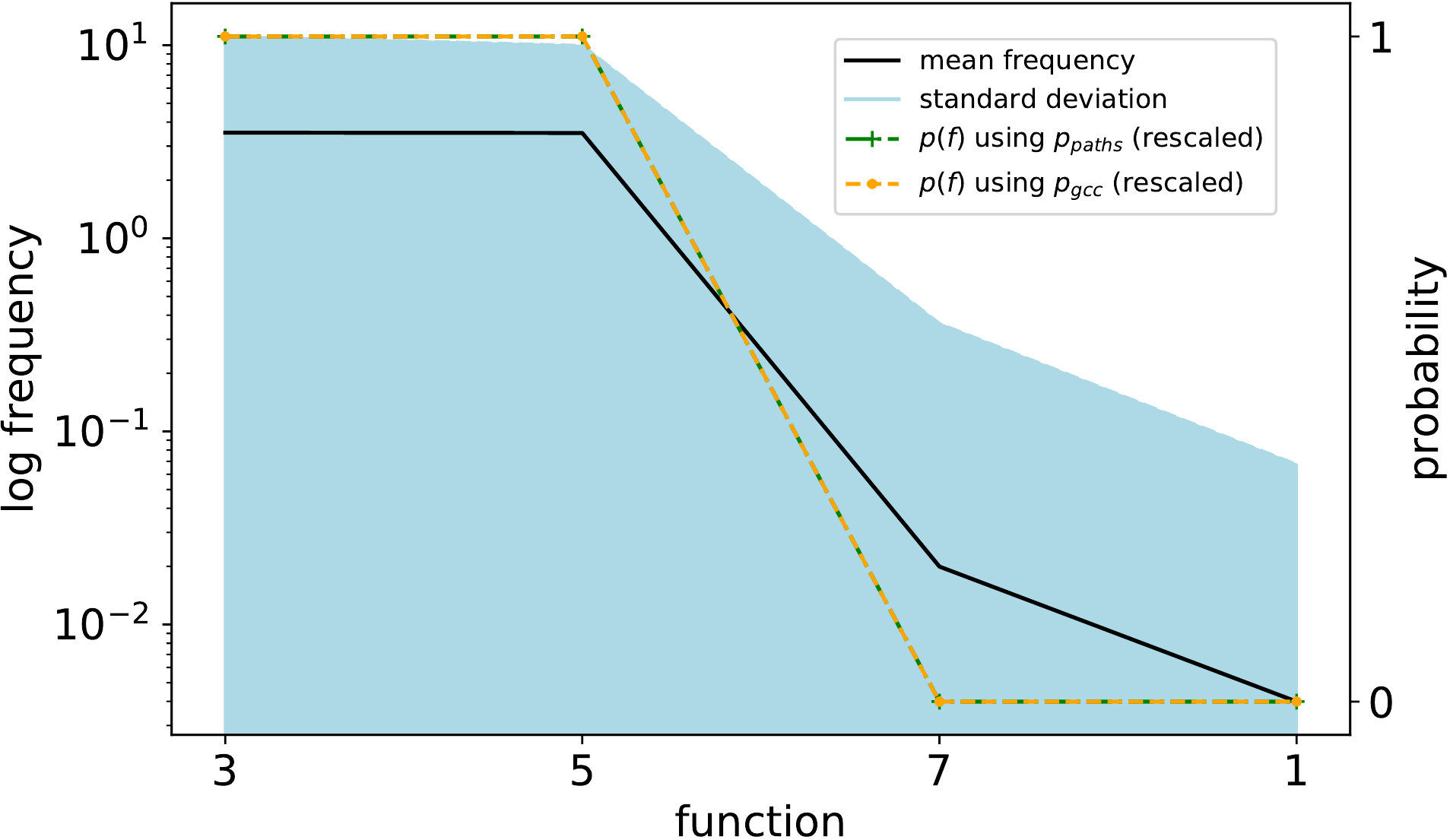}
         \caption{}
         \label{fig:rank_ordering_LTM}
     \end{subfigure}
    \hfill
     \begin{subfigure}[t]{0.47\textwidth}
     \vskip 0pt
     \includegraphics[scale =.43]{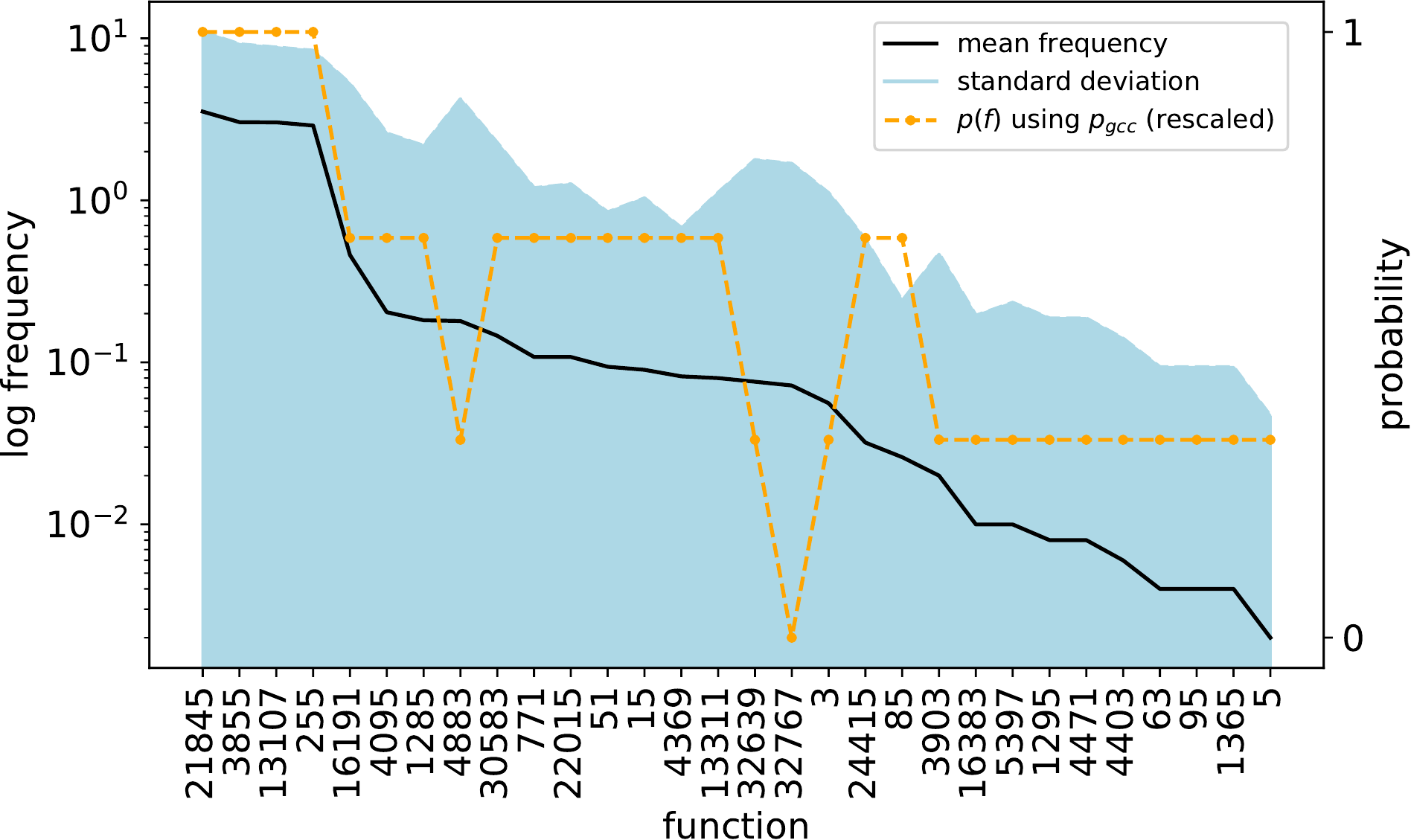}
         \caption{}
         \label{fig:rank_ordering_LTM_k_4}
     \end{subfigure}}
        \caption{Function frequency corresponds to probability of required paths in a rank-ordering.  (a) Logarithmic frequency of non-zero functions computed by the ensemble of LTM cascades for $N = 10000$ nodes, average degree $z = 4$ and $k = 2$ inputs, over 500 realizations reveals an apparent rank-ordering (solid line).  Mean frequency is proportional to path probabilities, having a Pearson correlation of $1.0$, both predicted by probabilities derived from logic motifs using $p_{path}$ ('+') (e.g. see (\ref{eq:prob_from_p_path})),  and complexity $p_{gcc}^{C(f) + 1}$ (large dot) (\ref{eq:p_propto_p_C+1}) (rescaled, overlaid, both dashed).  Thus (\ref{eq:p_propto_p_C+1}) also well-predicts (\ref{eq:prob_from_p_path}).  Frequency therefore varies inversely with decision tree complexity $C$ ('+'). (b) Rank-ordering is more evident for $k = 4$ inputs, appearing as a decreasing exponential with goodness of fit $r^2 = 0.88$. Again, $N = 10000$ and $z = 4$.  Here, Pearson correlation between $p(f)$ and mean frequency is $0.74$. Shaded regions are one standard deviation.  Probabilities have been centered and normalized.}
\end{figure}


A sufficiently large ALTM, by composing monotone decreasing functions (e.g. NAND, NOR), can undergo a cascade to compute any logical function on its nodes, forming a \textit{universal basis} \cite{savage1998models}.

\section{Statistics of attractors in the Boolean function space}

We experiment first on the LTM, to investigate the observed frequency of Boolean functions in simulation.  With a network having $N = 10000$ nodes, ensembled over 500 realizations, 
at mean degree $z = 4$ we observe that the frequency of functions is very skewed [Fig. \ref{fig:rank_ordering_LTM}].  Experiments for $k = 4$ inputs, again for $N = 10000$ nodes at mean degree $z = 4$, ensembled over $500$ realizations, also yield an approximate exponential decay of the rank ordering function [Fig. \ref{fig:rank_ordering_LTM_k_4}].

We investigate the skewed distribution of these functions by asking \textit{"What is the probability of obtaining the simplest network to compute each of these functions?"}. From Fig. \ref{fig:simplest_networks}, we can derive the probability of each monotone function.  For example, if there is no path from seed nodes $a$ and $b$ to some node $u$ we obtain $f_0$, thus

$$
p(f_0) \propto (1 - p_{path})^2,
$$

where $p_{path}$ is the probability of a path between two randomly chosen nodes.

The function $f_1$ requires paths from $a$ and $b$ to $u$, thus

\begin{equation}
    p(f_1) \propto p_{path}^2.
    \label{eq:prob_from_p_path}
\end{equation}

However, with percolation in mind, we observe that for large graphs, the probability of paths between $n$ nodes approaches the probability that all $n$ nodes belong to the giant connnected component (GCC) \cite{newman2018networks}. 

This gives us, again from Fig. \ref{fig:simplest_networks}, 
$$
p(f_1) \propto p_{path(A,B,u)} \propto p_{gcc}^3,
$$

where $p_{gcc}$ is the probability for a random node to belong to the GCC.

From \cite{newman2018networks}, we have the recursive relation 
\begin{equation}
p_{gcc} = v = 1 - e^{-zv},
\label{eq:p_gcc}    
\end{equation}
where $z$ is the mean degree.  

We subsequently observe that the number of required paths from seed nodes to node $u$, computing monotone function $f$, is equal to the \textit{decision tree complexity }($C$), the depth of the shortest decision tree to compute $f$.  In order for $u$ to decide the value of a seed node, the seed's perturbation information must be transmitted along a path to $u$.

Taking a Boolean function's Hamming cube representation, its decision tree complexity $C$ is complementary to the number of congruent axial reflections $R$ along each of its axes $D$ (details in supplemental information A.1) 
That is, if a Boolean function's Hamming cube is constant along an axis, it is independent of that axis, giving us
\begin{equation}
C = D - R.    
\label{eq:complexity_vs_symmetry}
\end{equation}

In other words, \textit{the number of paths a monotone Boolean function requires is exactly the number of axial reflection asymmetries of its Hamming cube.}

This allows us to relate function frequency to decision tree complexity.  Recall that the critical percolation threshold in an arbitrarily large Erdos-Renyi-Gilbert graph occurs at mean degree $z_c = 1$, a very small connectivity. Thus since $p \sim \frac{z_c}N$, $p_c \ll 1$.  Therefore, the network will be be tree-like, since the clustering coefficient $C_{\rm clus} \propto p$ \cite{newman2018networks}.  In a tree, the number of nodes is one more than the number of edges $N = |E| + 1$.  Thus, as $p \to p_c$,
\begin{equation}
    p(f) \propto p_{gcc}^{C(f) + 1}.
    \label{eq:p_propto_p_C+1}
\end{equation}

Indeed it appears that (\ref{eq:p_propto_p_C+1}) is highly correlated with the probabilities derived from logic motifs (\ref{eq:prob_from_p_path}), and that observed function frequency is proportional to (\ref{eq:p_propto_p_C+1}) as well [Fig. \ref{fig:rank_ordering_LTM}], having a Pearson correlation of approximately 1.0 for k = 2, and 0.74 for k = 4.  This also shows, due to (\ref{eq:p_propto_p_C+1}) an inverse rank ordering relation between frequency and decision-tree complexity, appearing as a decreasing exponential in frequency.    Given that, as mentioned in the introduction,  there is a increasing exponential distribution of decision tree complexity in the truth table of all Boolean functions, this result is especially surprising.

\subsection{Function Distribution with Antagonism}

\begin{figure}[ht]
     \centering
     \begin{subfigure}[t]{0.5\textwidth}
     \vskip 0pt
         \centering
\includegraphics[scale=.5]{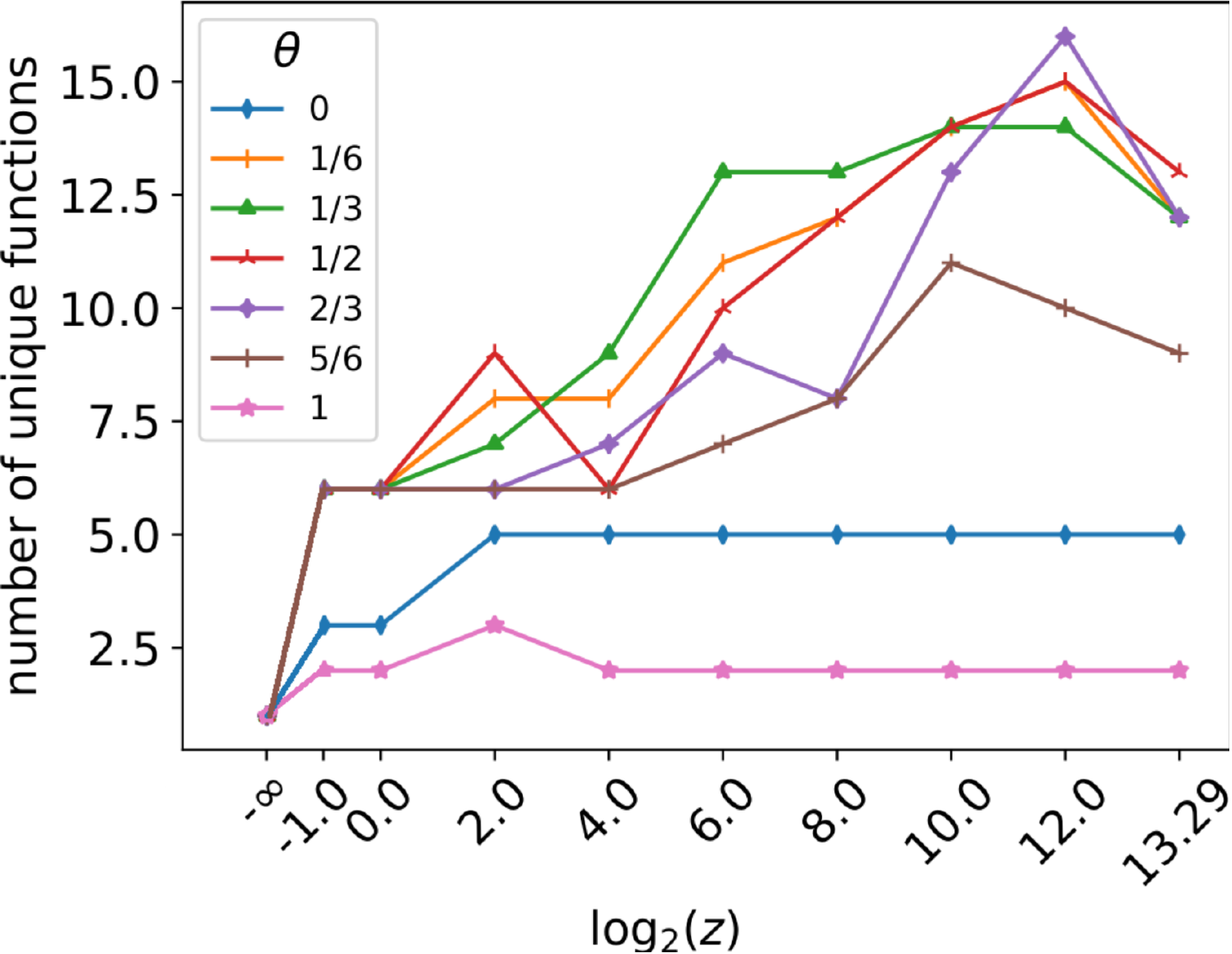}
\caption{}
         \label{fig:num_functions_ALTM}
     \end{subfigure}
     \hfill
     \begin{subfigure}[t]{0.47\textwidth}
     \vskip 0pt
         \centering
         \includegraphics[scale=.43]{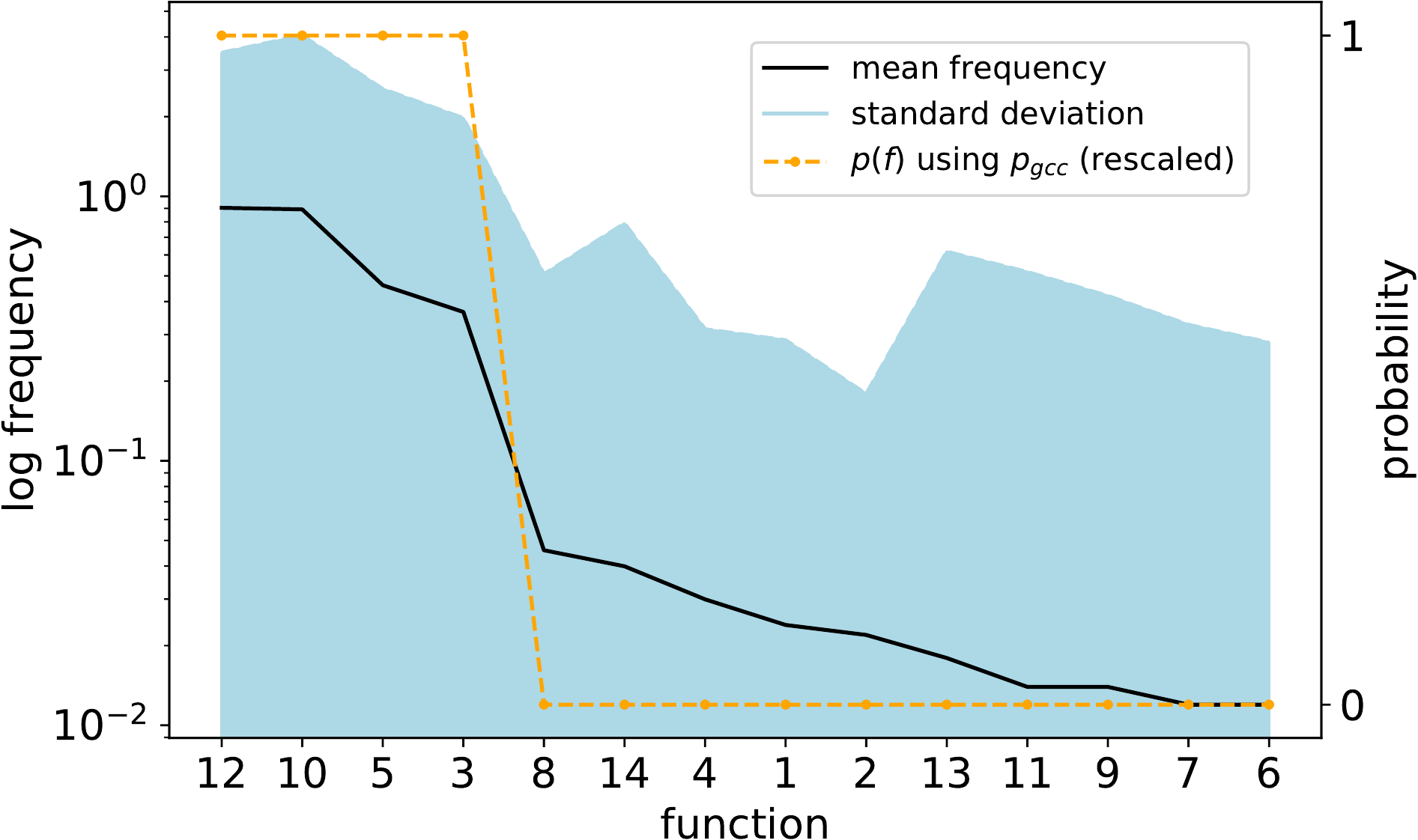}
         \caption{}
        \label{fig:rank_ordering_ALTM}
     \end{subfigure}
     
       \hfill
 
 \caption{Antagonism fraction ($\theta$) agrees with biology; non-monotone functions also predicted by path requirements.   (a) For networks with $N = 10000$ nodes and $k = 2$ inputs, over 500 realizations, varying the mean degree $z$ and fraction of antagonistic nodes $\theta \in \{0, \frac16, \frac26,... 1\}$, we observe that the mean number of unique functions per network is maximized over several orders of magnitude ($z \in [2^3, 2^{10}]$) by networks having a fraction of antagonistic nodes $\theta = \frac13$ (triangles), coinciding with other findings \cite{capano2015optimal}.
 (b)    At $\theta = \frac 13$ and $z = 2^6$, we again observe a skewed frequency, and a proportional relationship between function frequency and probability due to complexity (\ref{eq:p_propto_p_C+1}), having Pearson correlation of $0.91$. Shaded region is one standard deviation.  Probabilities have been centered and normalized. (Functions $f_0$ and $f_{15}$ have been removed, since in the ALTM they can occur outside of the GCC.) }
        
\end{figure}

A similar simulation, having $N = 10000$ nodes, $k = 2$ inputs, ensembled over 500 realizations in a range of mean degree values $z$ and fraction of antagonistic nodes $\theta \in \{0, \frac16, \frac26,... 1\}$, reveals a sudden increase in the number of unique non-zero functions vs. both $z$ and $\theta$ [Fig. \ref{fig:num_functions_ALTM}].   
The number of unique functions is maximized over several orders of magnitude near criticality, for $z \in [2^3, 2^{10}$], and $\theta = 1/3$.  Observing that antagonism and inhibition are interchangeable \cite{rojas2013neural} (Supplemental section A.2) 
, this lends support to optimal information processing around $30 \%$ inhibition, found in other research \cite{capano2015optimal}, and why this fraction of inhibitory neurons seems prevalent biologically. 

 \begin{figure}[ht]
         \centering
         \includegraphics[scale =0.7]{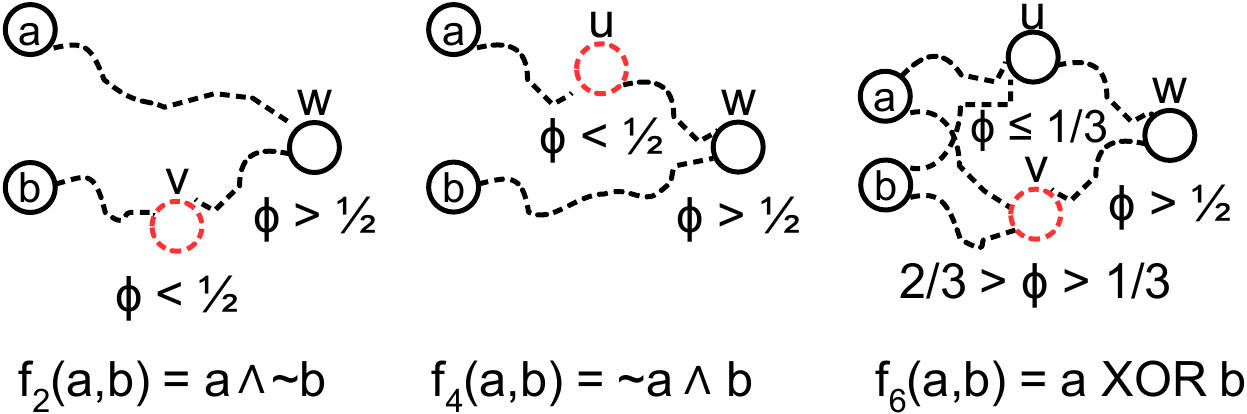}
         \caption{Motifs for non-monotone functions.  Simplest \textit{logic motifs} to compute non-monotone Boolean functions $\{f_2$, $f_4$, $f_6\}$ [Table \ref{truth_table2}] in the ALTM at random node $u$, on seed nodes $a, b$. Dashed lines represent paths, and dashed nodes are antagonistic.  Functions $f_{13}$, $f_{11}$, and $f_9$ are negations of these, respectively, so have very similar networks, negating each node.}
         \label{fig:non_monotone_motifs}
     \end{figure}
     
For this mix of LTM and ALTM nodes, we again observe a similar rank-ordering of functions, here at $z = 64, \theta = 1/3$, and that, as in the LTM, frequency is again proportional to probability derived from function complexity [Fig. \ref{fig:rank_ordering_ALTM}], having a Pearson correlation of 0.91.

We note, however, that (\ref{eq:complexity_vs_symmetry}) under-estimates the number of paths required for non-monotone functions.  For example, $f_6$ (XOR) requires 4 paths between 5 nodes, all of which must be in the GCC [Fig. \ref{fig:non_monotone_motifs}], so that $p(f_6) \propto p_{gcc}^5$.  However, this function's decision tree complexity $C = 2$, predicting by (\ref{eq:p_propto_p_C+1}) that $p(f_6) \propto p_{gcc}^3$.  Therefore a more informative complexity measure is needed for non-monotone functions.


\section{Discussion}

As indicated in the title, we see the main result of interest as the spontaneous emergence of complex logic functions in the minimally-constrained random threshold networks.  This then implies that many physical, biological, or other systems are able to perform such computation by ubiquitous avalanches or cascades.

We note that this result also begins to give us an explanation of the \textit{criticality hypothesis} vis-\`a-vis neuroscience \cite{massobrio2015criticality,hesse2014self,shew2013functional}.   That is, at the critical threshold, with the emergence of the giant component, the number of unique functions spontaneously increases.  Along with that comes an increase in the number of complex functions.  As neuronal networks need to compute integrative complex functions on sensory information, or on information passed between modular areas in the brain, the utility of this complexity is self-evident \cite{sejnowski1988computational}.  We note that in computational neuroscience, there is also discussion of the integration of information and complexity or consciousness \cite{tononi1994measure,lynn2019physics}. These motifs therefore give us a starting point for the relationship between structure and function as well.

Also, the present work connects to machine- or statistical-learning, where in classification, Boolean functions are computed on high-dimensional data.  Until now, however, despite their ubiquity in nature, neither criticality nor cascades have played a large role in machine learning as a design paradigm or analytical framework \cite{rojas2013neural}.  We see this as a large potential opportunity to improve deep learning methods.

The spontaneous emergence of complex computation is an example of a symmetry breaking phase transition, as the giant connected component (spanning cluster) comes into existence at the critical connectivity \cite{landau1937broken, anderson1972more}.  We conjecture that we are witnessing how \textit{complexity of functionality results from symmetry breaking in systems} \cite{anderson1972more}.  This complexity takes on a distribution that reflects a hierarchy in an exponential rank-ordering law.

We also see that, from a larger theoretical perspective, the confluence of cascades (percolation branching processes) and information processing by Boolean logic stands at the intersection between several very large and highly developed areas of research -- percolation- and computational automata-theory \cite{von1956probabilistic,christensen2005complexity}.

The specific mechanism of the logical automata realized by \textit{logic motifs} extends previous work about network motifs and their function, mainly in the genetic domain \cite{milo2002network}, into many other areas, again due to the ubiquity of cascades in threshold networks.

The observance of logic motifs as automata also allows us to change our perspective on network percolation.  In the past, we saw it perhaps only in terms of connected component size distribution.  Now, however, we may view these components as a \textit{zoo or library of functions}, available to the network by connection, much as importing a function occurs in programming languages.  We note that the scale invariance at criticality may exist at the Pareto-optimal point between complexity and diversity.  That is, there will be a small number of larger components computing complex functions, and a great number of very small, simple components having a large variety of thresholds.

\subsection{Future work}

In developing this work, we inevitably stumbled across an overwhelming number of ideas and directions that we can take.  We can only briefly list them.

We have seen above that other complexity measures could be found for non-monotone functions, to better predict their frequency in mixed LTM/ALTM networks.  We suspect that Boolean Fourier analysis would be fruitful here.  We also expect that, for larger inputs, these non-monotone functions will dominate the function space, and that the Hamming cube symmetries make it possible to write a partition function for them.  Along with this, it should be possible to predict more exact probabilities of functions, which depend on the occurrence of cascades being blocked, and of nodes inheriting their neighbors' complexity, among other factors.

We would also like to generalize these predictions to $k \gg 2$ inputs and much larger networks ($N \sim  10^9$ nodes), while understanding mechanisms and heuristics for learning by re-wiring in these large combinatorial spaces.
For example, we suspect that modularity develops as a network's capacity to extract complexity from inputs is exhausted.   We also suspect that function distribution can be understood in terms of multiple network density percolation thresholds, depending on function path requirements, more evident for larger inputs.

Furthermore, we intend to study the relation between function and network symmetry in the context of symmetry breaking.  We conjecture, for example, that there is a conservation law of complexity or information, meaning that what we call computation comes at the expense of lost information, rendering the network a kind of \textit{information engine} \cite{landauer1961irreversibility}, whose output is \textit{computation}, and that this lies at the heart of information creation.

Of course, it could also be fruitful to understand this work in terms of information processing, using measures such as transfer entropy, of increasing use in computational neuroscience and automata theory \cite{lizier2014framework}.  Along with this we see an opportunity to formalize the \textit{criticality hypothesis} in light of our results on computation.  In the hypothesis, avalanche criticality (the kind of percolation seen here) and so-called \textit{edge of chaos} are convolved qualitatively, by saying that information processing is optimized 'near criticality' \cite{beggs2008criticality,jensen2021critical}.

We would like to research the effects of geographic energy constraints and other network topologies, found in real-world systems, on the function phase space.  For example we conjecture that both modularity and layering will result from restricting geographic connection distance, with a result that complex functions appear at nodes on the surface (or interface) of networks, convenient for passing to subsequent networks.

Finally, although we have used the term \textit{computation} here, it would be useful to carefully study the linear threshold model as a computing machine, especially when re-wiring, investigating its Turing completeness, run-time, and related phenomena.

\section{Conclusion}

Here we have shown that the Linear Threshold Model computes a distribution of monotone Boolean logic functions on perturbation inputs at each node in its network, and that with the introduction of antagonism (inhibition), any function can be computed.  Notably, complex functions arise in an apparent exponentially decreasing rank-ordering due to their requirements for perturbation information from seed nodes, and these requirements correspond to their functional asymmetries.  These asymmetries can be used to obtain their probability exponent as a function of the probability of belonging to the network's giant connected component.  Finally, we observe that the number of unique functions computed by an LTM of mixed excitatory and antagonistic nodes is maximized near $1/3$ antagonism, over several orders of magnitude of connectivity, coinciding with other research.

\bibliographystyle{abbrv}
\bibliography{refs.bib}  

\providecommand{\noopsort}[1]{}\providecommand{\singleletter}[1]{#1}%
\begin{thebibliography}{10}

\bibitem{anderson1972more}
P.~W. Anderson.
\newblock More is different: broken symmetry and the nature of the hierarchical
  structure of science.
\newblock {\em Science}, 177(4047):393--396, 1972.

\bibitem{beggs2008criticality}
J.~M. Beggs.
\newblock The criticality hypothesis: how local cortical networks might
  optimize information processing.
\newblock {\em Philosophical Transactions of the Royal Society A: Mathematical,
  Physical and Engineering Sciences}, 366(1864):329--343, 2008.

\bibitem{ben2008farewell}
A.~Ben-Naim.
\newblock {\em A farewell to entropy: Statistical thermodynamics based on
  information: S}.
\newblock World Scientific, 2008.

\bibitem{brooks1988evolution}
D.~R. Brooks, E.~O. Wiley, and D.~Brooks.
\newblock {\em Evolution as entropy}.
\newblock University of Chicago Press Chicago, 1988.

\bibitem{capano2015optimal}
V.~Capano, H.~J. Herrmann, and L.~De~Arcangelis.
\newblock Optimal percentage of inhibitory synapses in multi-task learning.
\newblock {\em Scientific reports}, 5(1):1--5, 2015.

\bibitem{christensen2005complexity}
K.~Christensen and N.~R. Moloney.
\newblock {\em Complexity and criticality}, volume~1.
\newblock World Scientific Publishing Company, 2005.

\bibitem{conway1970game}
J.~Conway et~al.
\newblock The game of life.
\newblock {\em Scientific American}, 223(4):4, 1970.

\bibitem{dubbey2004mathematical}
J.~M. Dubbey and J.~M. Dubbey.
\newblock {\em The mathematical work of Charles Babbage}.
\newblock Cambridge University Press, 2004.

\bibitem{easley2010networks}
D.~Easley, J.~Kleinberg, et~al.
\newblock {\em Networks, crowds, and markets}, volume~8.
\newblock Cambridge university press Cambridge, 2010.

\bibitem{granovetter1978threshold}
M.~Granovetter.
\newblock Threshold models of collective behavior.
\newblock {\em American journal of sociology}, 83(6):1420--1443, 1978.

\bibitem{hesse2014self}
J.~Hesse and T.~Gross.
\newblock Self-organized criticality as a fundamental property of neural
  systems.
\newblock {\em Frontiers in systems neuroscience}, 8:166, 2014.

\bibitem{jalili2017information}
M.~Jalili and M.~Perc.
\newblock Information cascades in complex networks.
\newblock {\em Journal of Complex Networks}, 5(5):665--693, 2017.

\bibitem{jaynes1957information}
E.~T. Jaynes.
\newblock Information theory and statistical mechanics.
\newblock {\em Physical review}, 106(4):620, 1957.

\bibitem{jensen2021critical}
H.~J. Jensen.
\newblock What is critical about criticality: in praise of the correlation
  function.
\newblock {\em Journal of Physics: Complexity}, 2(3):032002, 2021.

\bibitem{kempe2003maximizing}
D.~Kempe, J.~Kleinberg, and {\'E}.~Tardos.
\newblock Maximizing the spread of influence through a social network.
\newblock In {\em Proceedings of the ninth ACM SIGKDD international conference
  on Knowledge discovery and data mining}, pages 137--146, 2003.

\bibitem{knuth2011information}
K.~H. Knuth.
\newblock Information physics: The new frontier.
\newblock In {\em AIP Conference Proceedings}, volume 1305, pages 3--19.
  American Institute of Physics, 2011.

\bibitem{landau1937broken}
L.~Landau.
\newblock Broken symmetry and phase transitions.
\newblock {\em Phys Z Sowjetunion, 1937, 11}, 26, 1937.

\bibitem{landauer1961irreversibility}
R.~Landauer.
\newblock Irreversibility and heat generation in the computing process.
\newblock {\em IBM journal of research and development}, 5(3):183--191, 1961.

\bibitem{lizier2014framework}
J.~T. Lizier, M.~Prokopenko, and A.~Y. Zomaya.
\newblock A framework for the local information dynamics of distributed
  computation in complex systems.
\newblock In {\em Guided self-organization: inception}, pages 115--158.
  Springer, 2014.

\bibitem{lloyd2010computational}
S.~Lloyd.
\newblock The computational universe.
\newblock {\em Information and the nature of reality: From physics to
  metaphysics}, pages 92--103, 2010.

\bibitem{lloyd2013universe}
S.~Lloyd.
\newblock The universe as quantum computer.
\newblock {\em A Computable Universe: Understanding and exploring Nature as
  computation}, pages 567--581, 2013.

\bibitem{lynn2019physics}
C.~W. Lynn and D.~S. Bassett.
\newblock The physics of brain network structure, function and control.
\newblock {\em Nature Reviews Physics}, 1(5):318--332, 2019.

\bibitem{massobrio2015criticality}
P.~Massobrio, L.~de~Arcangelis, V.~Pasquale, H.~J. Jensen, and D.~Plenz.
\newblock Criticality as a signature of healthy neural systems.
\newblock {\em Frontiers in systems neuroscience}, 9:22, 2015.

\bibitem{milo2002network}
R.~Milo, S.~Shen-Orr, S.~Itzkovitz, N.~Kashtan, D.~Chklovskii, and U.~Alon.
\newblock Network motifs: simple building blocks of complex networks.
\newblock {\em Science}, 298(5594):824--827, 2002.

\bibitem{newman2018networks}
M.~Newman.
\newblock {\em Networks}.
\newblock Oxford university press, 2018.

\bibitem{perseguers2010quantum}
S.~Perseguers, M.~Lewenstein, A.~Ac{\'\i}n, and J.~I. Cirac.
\newblock Quantum random networks.
\newblock {\em Nature Physics}, 6(7):539--543, 2010.

\bibitem{rojas2013neural}
R.~Rojas.
\newblock {\em Neural networks: a systematic introduction}.
\newblock Springer Science \& Business Media, 2013.

\bibitem{sakoda1971checkerboard}
J.~M. Sakoda.
\newblock The checkerboard model of social interaction.
\newblock {\em The Journal of Mathematical Sociology}, 1(1):119--132, 1971.

\bibitem{savage1998models}
J.~E. Savage.
\newblock {\em Models of computation}, volume 136.
\newblock Addison-Wesley Reading, MA, 1998.

\bibitem{schelling1971dynamic}
T.~C. Schelling.
\newblock Dynamic models of segregation.
\newblock {\em Journal of mathematical sociology}, 1(2):143--186, 1971.

\bibitem{schrodinger1992life}
R.~Schrodinger, E.~Schr{\"o}dinger, and E.~S. Dinger.
\newblock {\em What is life?: With mind and matter and autobiographical
  sketches}.
\newblock Cambridge university press, 1992.

\bibitem{sejnowski1988computational}
T.~J. Sejnowski, C.~Koch, and P.~S. Churchland.
\newblock Computational neuroscience.
\newblock {\em Science}, 241(4871):1299--1306, 1988.

\bibitem{shannon1941mathematical}
C.~E. Shannon.
\newblock Mathematical theory of the differential analyzer.
\newblock {\em Journal of Mathematics and Physics}, 20(1-4):337--354, 1941.

\bibitem{shannon1948mathematical}
C.~E. Shannon.
\newblock A mathematical theory of communication.
\newblock {\em The Bell system technical journal}, 27(3):379--423, 1948.

\bibitem{shew2013functional}
W.~L. Shew and D.~Plenz.
\newblock The functional benefits of criticality in the cortex.
\newblock {\em The neuroscientist}, 19(1):88--100, 2013.

\bibitem{tononi1994measure}
G.~Tononi, O.~Sporns, and G.~M. Edelman.
\newblock A measure for brain complexity: relating functional segregation and
  integration in the nervous system.
\newblock {\em Proceedings of the National Academy of Sciences},
  91(11):5033--5037, 1994.

\bibitem{turing1936computable}
A.~M. Turing et~al.
\newblock On computable numbers, with an application to the
  entscheidungsproblem.
\newblock {\em J. of Math}, 58(345-363):5, 1936.

\bibitem{von1951general}
J.~Von~Neumann.
\newblock The general and logical theory of automata, cerebral mechanisms in
  behavior. the hixon symposium.
\newblock {\em New York: John Wiley\&Sons}, 1951.

\bibitem{von1956probabilistic}
J.~Von~Neumann.
\newblock Probabilistic logics and the synthesis of reliable organisms from
  unreliable components.
\newblock {\em Automata Studies}, pages 43--98, 1956.

\bibitem{watts2002simple}
D.~Watts.
\newblock A simple model of global cascades on random networks.
\newblock {\em Proceedings of the National Academy of Sciences of the United
  States of America}, 99(9):5766--5771, 2002.

\bibitem{wheeler1992recent}
J.~A. Wheeler.
\newblock Recent thinking about the nature of the physical world: It from bit
  a.
\newblock {\em Annals of the New York Academy of Sciences}, 655(1):349--364,
  1992.

\bibitem{wheeler2018information}
J.~A. Wheeler.
\newblock {\em Information, physics, quantum: The search for links}.
\newblock CRC Press, 2018.

\bibitem{wilkerson2019universal}
G.~Wilkerson and S.~Moschoyiannis.
\newblock Universal boolean logic in cascading networks.
\newblock In {\em International conference on complex networks and their
  applications}, pages 601--611. Springer, 2019.

\bibitem{wolfram2002new}
S.~Wolfram.
\newblock {\em A new kind of science}, volume~5.
\newblock Wolfram media Champaign, IL, 2002.

\end{thebibliography}






\newpage
\section{Supplementary Information}

\subsection{Algorithm for determining Decision Tree Complexity}
\label{section:decision_tree_complexity}

\begin{figure}[h!]
    \centering
    \includegraphics[scale = 0.7]{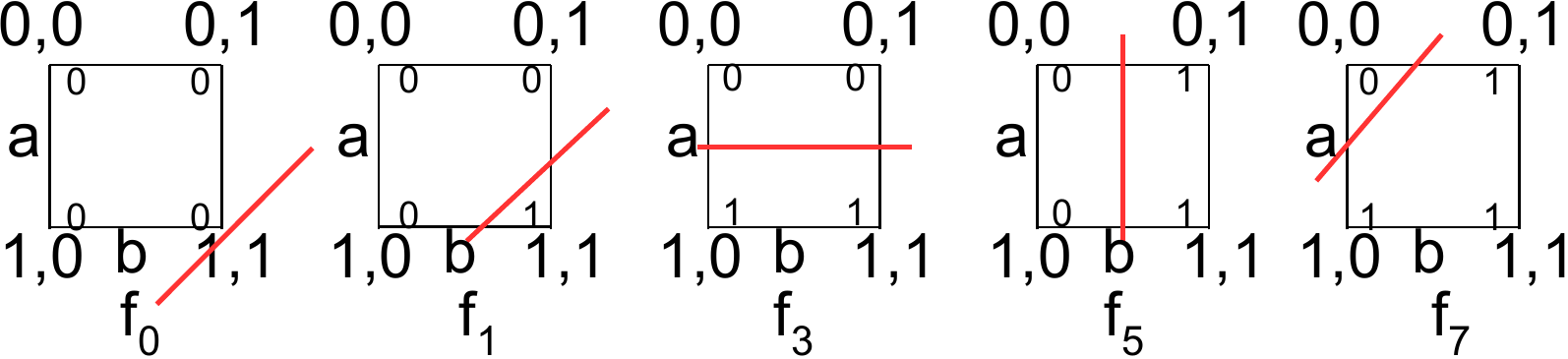}
    \caption{Hamming cube representations of LTM-computable monotone Boolean functions of two variables.  Line represents linear separator.  Note that for monotone functions, true values must be below or to the right of false values. }
    \label{fig:Hamming_cubes}
\end{figure}

\begin{algorithm}[h]
	\caption{Decision Tree Complexity using Hamming Cube reflections} 
	\label{alg:decision_tree_complexity}
	\begin{algorithmic}[1]
	\State $H \gets (x, f(x))$ \Comment{create labelled Hamming cube from inputs, outputs} 
	\State $R \gets 0$  \Comment number of congruent reflections
	
	\For{$d \in D$}  \Comment{for each dimension}
	
	\State $\Delta r = \left\{
            \begin{array}{ll}
                  1 & (H \equiv (H_d))  \\ 
                  0 & otherwise \\
            \end{array} \right. $
            
	\State $R \gets R + \Delta r$
	\EndFor
	
	\State $C \gets D - R$  \Comment Complexity = dimensionality - congruent reflections

    \end{algorithmic} 
\end{algorithm}
  
To determine a Boolean function's Decision Tree Complexity, we create the Hamming cube $H$ on the input values of the function.  The number of axes $D$ of $H$ is equal to the number of inputs ($k$).  We then label each corner of $H$ according to the function values $f$ [Fig. \ref{fig:Hamming_cubes}].  
    
    For each input variable $d$ of the cube's $D$ axes, we reflect the Hamming cube about that axis, obtaining the reflected Hamming cube $H_d$.  If $H \equiv H_d$, we add one to the reflection symmetry $R$.
    
    The Decision Tree Complexity $C$ is then the dimensionality $D$ minus the number of congruent reflections $R$ ($C = D - R$). The intuition is that if the Hamming cube of a particular function is congruent to an axial reflection, the function is independent of that axis.

    Thus, paths and their resulting cascades break symmetry and create complexity in the network, realized in the function order parameters.
\clearpage
\subsection{Interchangeability of Antagonism and Inhibition}
\label{section:interchangeability_antagonism_inhibition}

\begin{figure}[h!]
    \centering
    \includegraphics[scale =0.5]{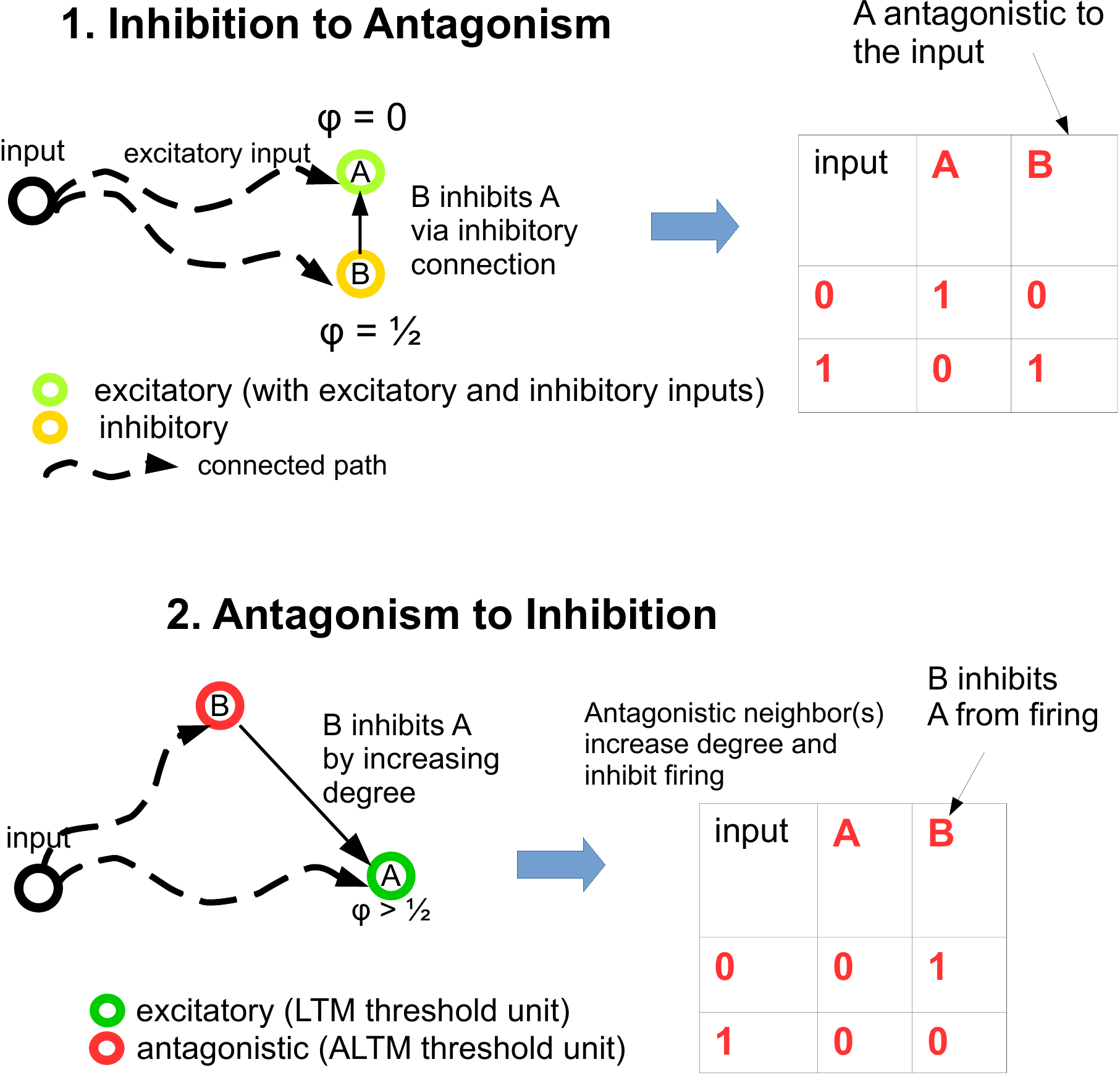}
    \caption{Simplest sub-networks to convert between inhibition and antagonism.  Both sub-networks have 2 internal nodes, which implies that there is a 1:1 ratio in the minimal number of nodes to perform either operation.}
    \label{fig:my_label}
\end{figure}

\end{document}